\documentstyle[sprocl]{article}
\def\Journal#1#2#3#4{{#1} {\bf #2}, #3 (#4)}

\def\PLB{{\em Phys. Lett.}  B}

\def\PRD{{\em Phys. Rev.} D}

\def\be{\begin{equation}}
\def\ee{\end{equation}}
\def\bea{\begin{eqnarray}}
\def\eea{\end{eqnarray}}

\begin{document}
\title{DECOHERENCE OF VACUUM FLUCTUATIONS IN COSMOLOGY}
\author{ MILAN MIJI\'C}
\address{Department of Physics and Astronomy,\\
California State University Los Angeles, CA 90032,\\
and, Institute for Physics, P.O. Box 522, 11001 Belgrade, Yugoslavia}
\maketitle\abstracts{A calculation of the one-mode occupation numbers for
vacuum fluctuations of massive fields in De Sitter space shows that
their use for the generation of classical density perturbations in inflationary
cosmology very much depends on their masses and conformal couplings. A new
mechanism for the decoherence of relatively massive fields, $2 < m^2/H_0^2
< 9/4$, has been identified. Similar analysis of the power law inflation
shows that production of adiabatic density perturbations may not take
place in models with power $p < 3$.}

\medskip
\noindent
Decoherence of vacuum fluctuations in inflationary cosmology \cite{Guth}
is at the foundation of the current paradigm for structure formation. General
arguments for classical behavior of fluctuations larger than the Hubble
radius \cite{rev} usually appeal to the existence of Gibbons-Hawking
temperature of De Sitter space, or to the rapid stretching of these modes
which collects them in a kind of infrared condensate. Attractive as they
are, such arguments were never very explicit, and leave some puzzles. For
instance, what about the decoherence of fluctuations of very massive fields?
Both mentioned arguments seem independent on mass, yet stochastic approach
to inflationary phase \cite{StarMeu} apparently shows that the random walk
of coarse-grained fields takes place only for fields with mass low compared to 
the Hubble scale.\cite{sdcgqf} It seems therefore appropriate to look for
a more explicit insight into this process.

\medskip
\noindent
Results reported here are based on the investigation of particle production 
of free fields in expanding universe.\cite{Mm} Similar lines of reasoning 
were pursued in Ref.'s 6-7. The key physical mechanism in all of these cases 
is rapid expansion of the spacetime background, so that classical behavior
emerges even in the case of free fields. This is in contrast with the
more frequent studies of the decoherence based on
the construction of a suitable reduced density matrix for the subsystem of
interest. For application of this later idea to fluctuations in inflationary
universe see Ref. 8.

\medskip
\noindent
Our method is to evaluate the one-mode occupation number when the wavelength
of the mode exceeds the Hubble radius. If the occupation number is large
we will have classical behavior.
For fluctuations of massive fields in spatially flat De Sitter space the
results are as follows.

\medskip
\noindent
(a) $m^2/H_0^2 < 2$, and $2 < m^2/H_0^2 \leq 9/4$. The one-mode occupation 
number at times after the mode crosses the Hubble radius grows as,

\be
n(z) \sim \left ( \frac {\lambda_{phys}(z)}{H_0^{-1}} \right )^{2 \nu}~~.
\ee

\noindent
Here, $z \equiv - k\eta$ is a new time variable ($k$ is fixed). $\lambda_{phys}
\equiv S(\eta) k^{-1}$ is the physical wavelength, and parameter $\nu$ 
measures the mass in Hubble units: $\nu^2 \equiv 9/4 - m^2/H_0^2$. Since
the occupation number diverges each mode forms a classical condensate. 
For $m^2/H_0^2 < 2$ this happens because modes outside the Hubble radius 
roll along the upside-down potential.\cite{GuthPi} For 
$2 < m^2/H_0^2 \leq 9/4$ the potential is
upside-right at all times, including the times later than the Hubble crossing
time, but the occupation number still diverges due to the
dominance of just one mode (as opposed to simultaneous presence of both) as
the amplitude of this upside-right oscillator settles to the minimum. In both 
cases the same criteria for classical behavior are satisfied. 

\medskip
\noindent
(b) $m^2/H_0^2 =2$. Minimally coupled field with this mass is equivalent to 
conformally coupled massless field, so there is no particle production 
whatsoever.

\medskip
\noindent
(c) $m^2/H_0^2 > 9/4$. In this case the late time behavior of $n(z)$ is 
purely oscillatory,
\begin{equation}
n(z) = A_0 ~ +~
 A_c \cos \left [ 2 |\nu| \log (z/2) - 2 \Phi_{\Gamma} \right ]
~+~
A_s \sin \left [ 2 |\nu| \log (z/2) - 2 \Phi_{\Gamma} \right ] ~~.
\label {nosc}
\end{equation}

\noindent
$\Phi_{\Gamma}$ is phase fixed by the mass.
The potential is upside right in this case, but both oscillatory modes must 
be kept. This leads merely to oscillations in some finite particle number,
not to its divergence. One can show that the amplitude is bounded as
$n(z) \leq f \equiv A_0 + ( A_c + A_s)^{1/2} \ll 1$, for 
$m^2/H_0^2 - 9/4 \geq 1$. 

\medskip
\noindent
(d) {\it Non-minimal coupling.} The effect of adding an $\xi R \phi^2/2$ term is particularly simple in case
of a flat De Sitter background: in all the expressions above one should
replace $m^2$ with $m^2 + 12 \xi H_0^2$. The behavior of the occupation
number as parameterized by different values of this quantity is the same as 
before. The basic physics of the phenomena is unchanged.

\medskip
\noindent
For quantum fluctuations of inflaton that drives the power law inflation
$a \sim t^p, p > 1$, one finds that the convenient characteristic parameter 
is the power of expansion $p$.
The corresponding characteristic values are as follows: for 
$m^2/H_0^2 = 9/4$ in De Sitter case we have $p = 3$ in case of power 
law inflation, while $m^2/H_0^2 = 2$ corresponds to
$p =p_+ \equiv (7 + \sqrt {33})/4 \approx 3.186$. The equation of motion, 
boundary conditions and solutions are the same. From the analysis of the De 
Sitter case one deduces the following decoherence properties of vacuum 
fluctuations in power law inflation:

\medskip
\noindent
(a) $p > p_+$, and $3 \leq p < p_+$. The occupation number diverges as the
wavelength exceeds the Hubble radius. For $p > p_+$ the oscillators are 
upside-down, and the argument of Guth and Pi \cite{GuthPi} applies.
For $3 \leq p < p_+$  the oscillators are always upside-right but the 
occupation number nevertheless diverges due to the dominance of just one mode.

\medskip
\noindent
(b) $p = p_+$. Fluctuations in this case are the same as that for a massless
minimally coupled field, and there is no particle production.

\medskip
\noindent
(c) $p < 3$. The oscillators are upside-right, but classical solutions are
oscillatory and both modes must be kept. The occupation number is finite,
but oscillatory, and always smaller then unity.

\medskip
\noindent
Therefore, only the power law inflation of type (a)
has classical adiabatic perturbations. These results may also have some
consequence for the suspected increase of power towards large scales in
models of extended inflation.

\medskip
\noindent
To conclude, we find the traditional particle production to be a simple
and explicit
mechanism, sufficient to describe evolution of vacuum fluctuations into
classical perturbations. The same method may be applied to study of the
decoherence in any Robertson-Walker spacetime.

\end{document}